\documentclass[final,1p,times]{elsarticle} 
\usepackage{graphicx} 
\usepackage{amssymb} 
\usepackage{amsthm} 
\usepackage{lineno} 

\journal{Nuclear Physics A} 
\begin{document} 

\begin{frontmatter} 


\title{SPS energy scan results and physics prospects at FAIR}

\author{C.~H\"ohne}

\address{GSI Helmholtzzentrum f\"ur Schwerionenforschung Darmstadt GmbH,
Germany}

\begin{abstract} 
Experimental studies of nucleus-nucleus collisions in the whole
SPS energy range are reviewed. Selected topics such as statistical
properties of the hadronic phase, strangeness production,
fluctuations and correlations are discussed with regard to
information on the onset of deconfinement and the critical point
of strongly interacting matter. In spite of the very interesting
results obtained in particular at the low SPS energies, additional
data including rare probes such as charmed particles and
di-leptons are required for a precise understanding of the
underlying physics. An outlook about prospects and capabilities of
upcoming experiments in this interesting energy region at RHIC,
SPS, and in particular with CBM at FAIR, is given.
\end{abstract} 

\end{frontmatter} 



\section{Energy scan at the CERN-SPS}

In 1997, with the first results from central Pb+Pb collisions at
the top-SPS energy of 158~$A$GeV beam energy coming in, NA49
\cite{na49_proposal} proposed an energy scan in order to search
for signs for an onset of deconfinement for which early
predictions have been made \cite{marek1}. After a SPS test in
1998, Pb-ions at 4 different beam energies below the top energy
were provided by the CERN-SPS: 5 weeks of 40 $A$GeV beam energy
(1999), 5 days of 80 $A$GeV beam energy (2000), and 7 days each of
20 $A$GeV and 30 $A$GeV beam energy (2002). Table \ref{sps_exp}
lists these beam energies, the participating SPS experiments and
their main observables for which the energy dependence was
studied. NA45/ CERES, NA49 and NA57 were the main participants;
NA50 and NA60 basically concentrated on the highest energy.

 \begin{table}[b]
 \begin{tabular}{|c|c|c|}
 \hline
 beam energy [$A$GeV] & experiments & main observables (energy
 dep.) \\
 \hline
 40, 80, 158 & NA45/ CERES & di-electrons, HBT, $\langle p_{t}\rangle$-fluctuations\\
  20, 30, 40, 80, 158 & NA49 & hadron production, strangeness,
  fluctuations \\
  & & HBT, flow, light fragments, correlations\\
  40, 158 & NA57 & strange hyperons\\
  \hline
 \end{tabular}
 \caption{Overview on participation of the SPS-Experiments at different beam energies and their main observables studied in dependence on energy.}
\label{sps_exp}
 \end{table}

\begin{figure}[ht]
\includegraphics[height=0.29\textheight]{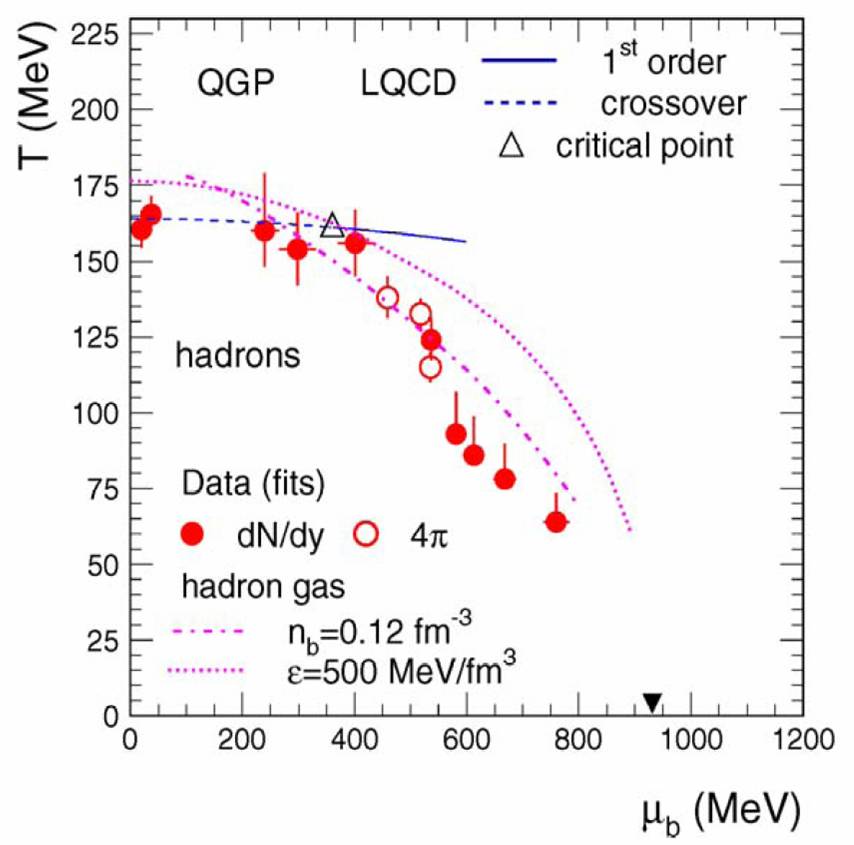}
\hfill
\includegraphics[height=0.29\textheight]{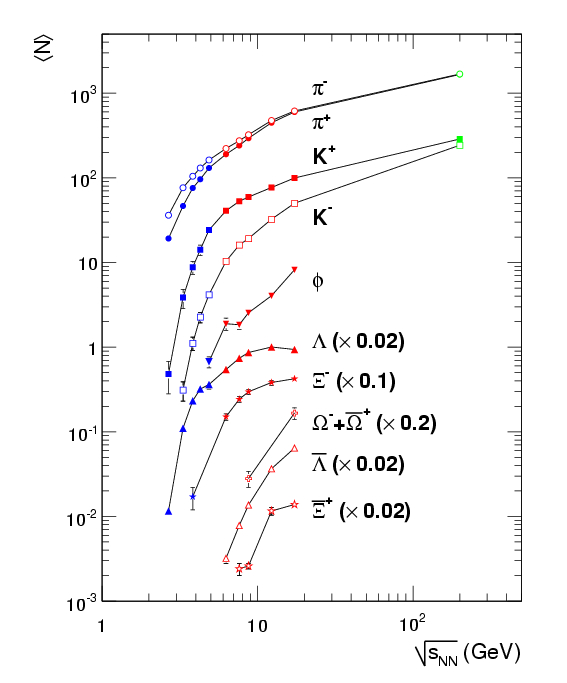}
 \caption{Left frame: Phase diagram \cite{pbm_old} with data points from central $A+A$ collisions and lattice QCD calculations from
 \cite{lattice}. Right frame:
 Measured 4$\pi$ multiplicities in central Au+Au/ Pb+Pb collisions at AGS (blue), SPS (red), and RHIC (green) versus center-of-mass energy \cite{4pi_yields_all}. }
\label{hadrons}
\end{figure}

The main objective of this systematic study of hadron production
was the search for the lowest energy which is sufficient to create
a partonic system in central Pb+Pb collisions, the so-called
''onset of deconfinement''. The parameters ($T, \mu_{B}$)
extracted from a statistical model analysis of the observed
hadronic freeze-out state lie in a region of the QCD phase diagram
of strongly interacting matter (fig.~\ref{hadrons}, left), for
which recent lattice calculations predict a first order phase
transition and a critical endpoint \cite{lattice}. Indications for
these features were searched for in the experiments. Finally, the
aim of course is to quantitatively characterize the created dense,
strongly interacting matter for which penetrating probes are most
useful. Recently, new exotic phases predicted at high baryon
density \cite{mclerran} have become topics of interest. In the
following, results from investigations of the energy dependence at
the SPS will be summarized. For more details and figures the
reader is referred to the quoted papers.

\subsection{Hadron production and search for the onset of deconfinement}

Yields and kinematical distributions of a large variety of hadrons
have been collected in dependence on beam energy
\cite{na49_kpi_energy,na49_energy,ceres_dileptons,na57} (see
fig.~\ref{hadrons}, right). However, data for more rare probes
such as multi-strange hadrons and di-leptons are missing for
center of mass energies $\sqrt{s_{NN}}<(8-9)$~GeV, i.e. in the AGS
and lowest SPS energy domain. They partially start becoming
available for $\sqrt{s_{NN}}<2$~GeV from the FOPI and HADES
experiments at GSI. Charmonium production in $A+A$ collisions has
so far been only measured at top-SPS energy
($\sqrt{s_{NN}}=17.3$~GeV) and above. Globally, particle
multiplicities rise with energy as expected. Interesting
structures, however, become visible in particle ratios, kinetic
properties and particle correlations. Together they give strong
evidence that a partonic phase is reached in A+A collisions from
about 30~$A$GeV beam energy upwards ($\sqrt{s_{NN}}=7.6$~GeV).

\begin{figure}[ht]
\centering
\includegraphics[height=0.245\textheight]{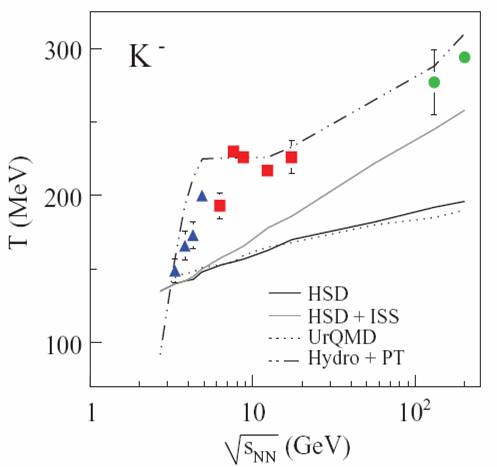}
\hfill
\includegraphics[height=0.245\textheight]{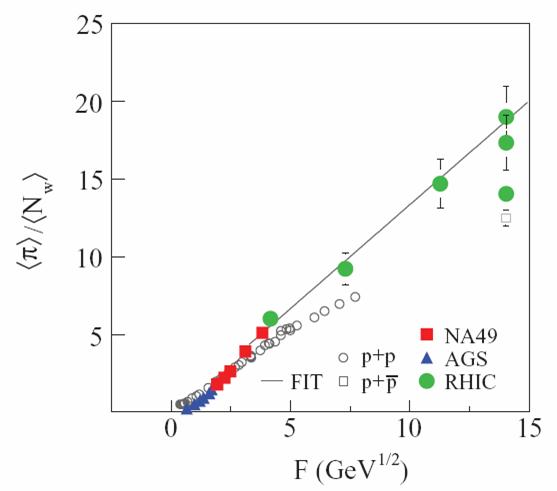}
 \caption{Inverse slope of $K^{-}$ $m_{t}$-spectra (left) and $\langle \pi \rangle/\langle
N_{w} \rangle$ ratio (right) in central Pb+Pb/ Au+Au collisions
versus energy with data from AGS (blue), SPS (red) and RHIC
(green) experiments \cite{na49_kpi_energy}. } \label{fig3}
\end{figure}

A striking feature is the steady behaviour of mean transverse
masses $\langle m_{t} \rangle - m_{0}$ in the SPS energy range
(fig.~\ref{fig3}, left) while for lower and higher energies a rise
is clearly seen \cite{na49_kpi_energy,na49_energy}, at least if
data from the AGS experiments are available to compare with. This
''steplike'' behaviour can be understood in terms of the creation
of a mixed hadronic and partonic phase in $A+A$ collisions at
these energies. Pressure and temperature become independent of the
energy density resulting in rather constant $\langle m_{t} \rangle
- m_{0}$ values \cite{step}. More recent hydrodynamical
calculations also support the importance of a (strong) 1st order
phase transition in order to explain the data \cite{hydro} which
would be difficult to describe otherwise.

The observed increase of the measured $\langle \pi \rangle/\langle
N_{w} \rangle$ ratio in central $A+A$ collisions compared to p+p
interactions \cite{na49_kpi_energy} (fig.~\ref{fig3}, right) can
be attributed to an increase in the number of degrees of freedom
due to deconfinement \cite{marek}. When extracting the sound
velocity from the width in rapidity of the $\pi$ spectra a minimum
at 30 $A$GeV beam energy has been observed \cite{sound}. Such a
minimum would be expected if the softest point of the nuclear
equation of state is reached, which might happen just at the onset
of deconfinement.

\begin{figure}[ht]
\centering
\includegraphics[height=0.24\textheight]{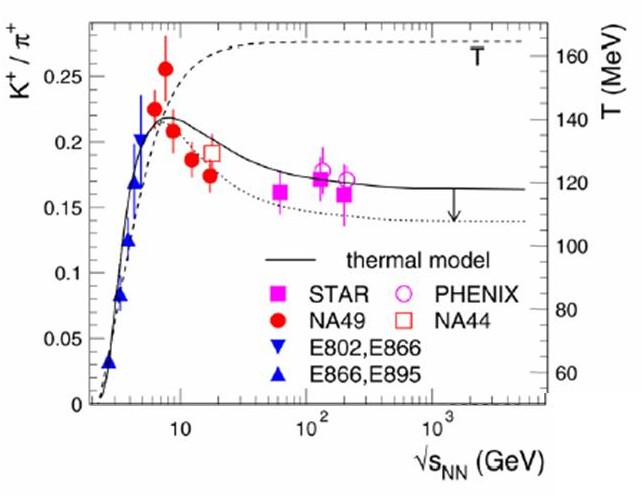}
\hfill
\includegraphics[height=0.24\textheight]{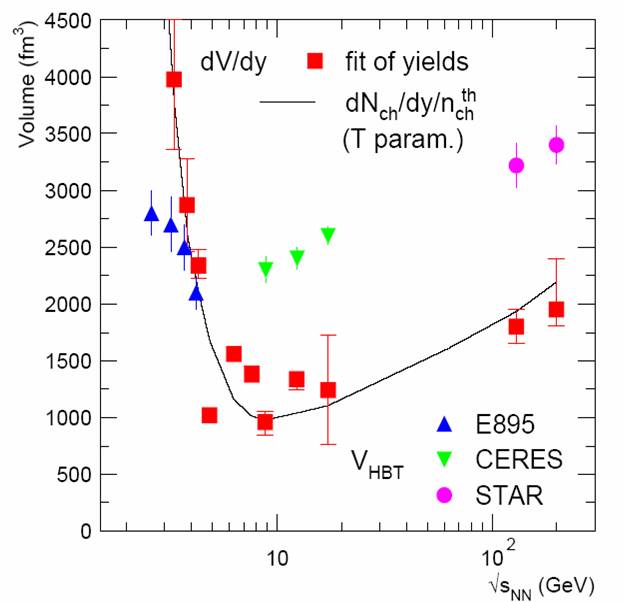}
 \caption{Left frame: $\langle K^{+} \rangle/\langle \pi^{+} \rangle$ ratio versus energy. As solid line the new thermal model calculation \cite{pbm_new}
 is shown, the dotted line shows this calculation including the full Hagedorn mass spectrum, the dashed line indicates the saturating temperature (right scale).
 Right frame: Volume versus energy as extracted from the thermal model (red squares) and the Bose-Einstein correlations
 of pions (dots, triangles) \cite{volume}. }
\label{fig4}
\end{figure}

One of the most discussed observations is probably the maximum in
relative strangeness production at, again, 30 $A$GeV beam energy
\cite{na49_kpi_energy}. So far, it could only be described and was
indeed predicted in a statistical model that explicitly included
an early partonic phase in central $A+A$ collisions from 30 $A$GeV
beam energy upwards \cite{marek}. Equilibrium hadron gas models
\cite{shm} and microscopic transport calculations \cite{hsd1}
always have difficulties in describing the rather pronounced
maximum, in particularly seen in the $\langle K^{+}\rangle/\langle
\pi^{+} \rangle$ ratio. In this context, a new statistical model
calculation including an improved hadron resonance mass spectrum,
in particular resonances with higher masses and the $\sigma$
meson, is noteworthy \cite{pbm_new}. Due to enhanced $\pi$
production from e.g.\ higher lying K$^{*}$ resonances the maximum
in the relative K$^{+}$ production is sharpened. When using an
exponential extrapolation of the measured hadron mass spectrum to
even higher masses and a Hagedorn temperature parameter $T_{H}\sim
200$ MeV, this effect is enhanced and data and model come into
better agreement (fig.\ \ref{fig4}, left). The maximum reached at
about $\sqrt{s_{NN}}\sim 8$ GeV is connected to the fact that the
baryochemical potential decreases strongly while the chemical
freeze-out temperature saturates at $\sim$ 164 MeV for high beam
energies. This implicitly points towards reaching the phase
boundary of partonic and hadronic matter in the SPS energy domain:
Although the energy density in central $A+A$ collisions still
increases from SPS to RHIC energies the saturation of the
freeze-out temperature can be seen as evidence that the additional
energy goes into the transition from hadronic to partonic matter
and/or into heating the partonic phase and the equilibrated hadron
gas is established at freeze-out at the phase boundary. In this
framework also the apparent chemical equilibrium of strangeness
could be explained \cite{wetterich}. It therefore would be
extremely interesting to test whether at low SPS and AGS energies
strangeness is equilibrated as well. Unfortunately precise data on
multi-strange hyperons are missing.

In this context it is interesting to note that the volume at
chemical freeze-out from statistical models and the correlation
(kinetic freeze-out) volume determined from Bose-Einstein
correlations of pions both show a minimum at SPS energies
\cite{hbt,volume} (fig.~\ref{fig4}, right). The latter is not
confirmed by NA49 \cite{hbt} which has an interesting consequence:
In \cite{hbt_ceres1} the freeze-out condition was formulated in
terms of a constant pion mean free path of appr.~1~fm. This holds
astonishingly well for all energies. However, a recent analysis
including also the NA49 data indicates the possibility of a
maximum at SPS energies \cite{hbt_ceres2}. This could be connected
to a slower expansion of the system resulting in more diffusive
character as would be expected near the softest point in the
equation of state. Unfortunately, identified particle elliptic
flow measurements \cite{flow} which could support this
interpretation are not conclusive. In \cite{akkelin} Bose-Einstein
correlations of pions in combination with their $p_{t}$ spectra
have been interpreted as being due to a phase transition into
partonic matter: A plateau in the averaged phase-space density of
pions is reached at SPS because the energy density is transformed
into new degrees of freedom and the pion density seen in
experiment stems from later freeze-out.

Complementary information is gained investigating the centrality
or system-size dependence of hadron production at different
energies. Of particular interest is e.g.~how the relative
strange\-ness production changes from p+p to central Pb+Pb for
energies below, at, and above $\sqrt{s_{NN}}\sim 8$~GeV where the
above discussed maximum is observed. The data, unfortunately, are
scarce. Available data on hyperon production
\cite{na57,NA49_centrality} at 40 and 158 $A$GeV beam energy can
be explained with a statistical model ansatz if the relevant
volume is calculated taking geometrical considerations into
account \cite{percolation}. Whether the changing shape of the
centrality dependence of relative strangeness production from a
rather linear rise at low energies \cite{ags_sis} to a fast rise
for small systems with subsequent saturation can also be
understood as being connected with the phase boundary remains open
until more data become available.

\subsection{Search for the critical point}

Signs for approaching the critical point in the QCD phase diagram
were searched for in event-by-event fluctuations of hadronic
observables. In case of a second order phase transition they could
be due to density fluctuations resulting in critical opalescence.
They might also appear due to a coexistence region of hadronic and
partonic matter in the course of a first order phase transition.
Fluctuation studies have been performed for multiplicity
fluctuations, net electric charge fluctuations, $\langle p_{t}
\rangle$-fluctuations and particle ratio fluctuations
\cite{fluctuations}. The observations are difficult to extract
experimentally as many effects such as centrality definition,
particle identification or phase space acceptance play a role. The
results are not conclusive and many effects are under discussion
for their interpretation \cite{theory_fluc}. In \cite{katia} a
discussion of recent results together with model predictions for a
critical point can be found.


\subsection{Study of the initial phase with
penetrating probes}

Observables sensitive to the high density or the partonic nature
of the medium created in central Pb+Pb collisions at SPS should
ideally be studied below, at, and above $\sqrt{s_{NN}}\sim 8$~GeV
where the onset of deconfinement is probably observed. However,
also here the data are scarce. Data on elliptic flow of hadrons
which can be related to the initial pressure and equation of state
are not conclusive \cite{flow}, and charmonia and direct photons
are not measured below top-SPS energy. Recent measurements
investigate two particle angular correlations at high $p_{t}$.
Indications for (mini-)jets at top SPS energies have been found
where the away side clearly shows medium, not jet behaviour
\cite{jetsCERES}. First data on the full SPS energy dependence
show that the near side correlation disappears at low energies
which might be due to momentum conservation. Interestingly, the
away side shows a plateau at all energies \cite{jetsNA49}.

Di-leptons as penetrating particles directly probe the high
density phase of $A+A$ collisions. An enhancement in the mass
range of (0.2-0.6)~GeV/c$^{2}$ above the known hadronic sources
has been found at 40 and 158 $A$GeV beam energy
\cite{ceres_dileptons}. Interestingly the enhancement is about a
factor 2 larger at the lower energy. This observation has been
interpreted as being a sign of the importance of baryon density
for the explanation of the enhanced di-lepton yield \cite{rapp}.
Although at SPS energies the $\rho$-meson yield is mainly
dominated by $\pi-\pi$ scattering, the contribution at low
invariant masses of the di-lepton pair is due to the coupling to
baryonic resonances. These processes can be studied at very low
energies by the HADES spectrometer \cite{hades}. Due to the
combination of p+p and n+p data as experimental reference spectrum
the effect of baryon density in $A+A$ collisions can be
investigated. Light collision systems such as C+C are still
explainable by this reference spectrum extracted from the p+p and
p+n data. However, new data from Ar+KCl collisions at 1.76 $A$GeV
beam energy show an enhanced di-electron yield which has to be
attributed to the baryon density in the medium \cite{hades}.

\section{Future explorations}


Strong evidence has been found from the SPS experiments that a
partonic phase is produced in central Pb+Pb collisions from about
30 $A$GeV beam energy onwards. This conclusion is based on
distinct changes in the behaviour of particle yields, spectra and
correlations of emitted particles which have been discussed above.
The evidence for the softest point of the equation of state is not
yet conclusive. The critical point has not been found so far. The
investigation of medium probes such as jets and di-leptons
suggests strong influence of the initially created strongly
interacting matter, but more data are necessary for quantitative
evaluation. In summary, the exploration of the QCD phase diagram
at the SPS has shown that the investigated energy range exhibits
very interesting features. The interest in this region is
strengthened by lattice QCD calculations which predict that a
first order phase transition and a critical endpoint are located
in this region. For a better and quantitative characterization of
the created media better data are needed for energies from appr.\
(2-60) $A$GeV beam energy: higher statistics in order to, e.g.,
also address elliptic flow with identified particles, more
systematic investigations, and new observables such as rare probes
(di-leptons, charm).

 \begin{table}[t]
 \begin{tabular}{|c|c|c|c|c|}
 \hline
 \multicolumn{2}{|c|}{energy range} & appr.\ $\mu_{B}$ range & experiments & starting time \\
 $\sqrt{s_{NN}}$ [GeV] & $E_{lab}$ [$A$GeV] & [MeV] & &\\
 \hline
 5 (62.4) - 200  & & $< 540$ & STAR, PHENIX at RHIC & 2010\\
 4.5 - 17.3 & 10 - 158 &  220 - 580  & NA61/SHINE at SPS & 2010 \\
 3 - 9 & & 360 - 700& MPD at NICA & 2014\\
 2.3 - 8.2 & 2 - 35 & 380 - 780& HADES, CBM at FAIR & 2016 \\
  \hline
 \end{tabular}
 \caption{Overview on future experiments and the foreseen beam energy ranges (for Au ions) \cite{na61,cbm,rhic,nica}. The approximate range in $\mu_{B}$ is taken from \cite{pbm_new}.}
\label{future}
 \end{table}

Table \ref{future} lists all future experimental programs which
are planned in order to address these questions and search for the
structures in the QCD phase diagram (see also
\cite{na61,cbm,rhic,nica}). All listed experiments will be able to
investigate hadron production including multi-strange hyperons,
fluctuations and correlations in more detail than has been done so
far. Rare probes such as di-leptons and charm production will only
become accessible at FAIR where sufficiently high beam intensities
of up to $10^{9}$ ions/s will be provided.

In the RHIC beam energy scan the STAR and PHENIX experiments will
concentrate on the evolution of medium properties which have been
measured in great detail at the top RHIC energies. A good
signature to look for is e.g.\ the ''turn-off' of established
effects such as the quark-number scaling of identified particle
elliptic flow or the suppression of hadrons with high $p_{t}$. The
critical point and a first order phase transition will be searched
for in fluctuations for which in particular the STAR detector with
its large TPC is well suited. The successor of NA49, the
NA61/SHINE experiment at the SPS will focus its search on
event-by-event fluctuations, to be measured systematically with
different energies and system sizes in order to find the critical
point. The MPD detector at the new accelerator project NICA at
Dubna plans to systematically study multi-strange hyperons and
flow.

\subsection{Physics prospects with CBM at FAIR}

The Compressed Baryonic Matter (CBM) experiment at FAIR will
measure hadron and lepton production in p+p, p+A and A+A
collisions from (8-35) $A$GeV beam energy (Au ions, up to 45
$A$GeV for ions with $Z/A=0.5$); the upgraded HADES detector will
cover the range from (2-8)~$A$GeV beam energy. Main objectives of
CBM are the search for signatures of a phase transition and the
critical point by a careful scan of the excitation functions of
hadron production, correlations, and fluctuations including rare
probes such as charm production and di-leptons. The aim is to also
quantify the properties of the created media, to establish the
equation of state at high baryon densities and to search for the
onset of chiral symmetry restoration.

Due to the high availability of the beam (order of 10 weeks
integrated beam time per year) and beam intensities up to $10^{9}$
ions/s systematic investigations with high statistics are
foreseen. For the physics prospects of CBM the following 3 cases
for a usage of 10 weeks Au-beam at 25 $A$GeV beam energy may serve
as an illustration: Minimum bias Au+Au collisions at 25 $A$GeV
beam energy without trigger can be recorded with 25 kHz
interaction rate. This yields basically ''unlimited'' statistics
of bulk observables (order of $10^{10-11}$ Kaons and $\Lambda$),
low-mass di-electrons with high statistics ($\sim 10^{6}$ $\rho$-,
$\omega$-, and $\phi$-mesons each), and multistrange hyperons with
high statistics ($10^{8} \Xi$, $10^{6} \Omega$). This will allow
for the measurement of yields, spectra, flow, correlations and
fluctuations in order to study the scaling behaviour of flow, the
equation of state or strangeness equilibrium. If an open charm
trigger is used, CBM will run with 100~kHz interaction rate which
allows to collect on the order of $10^{4}$ open charm hadrons.
With a charmonium trigger, the maximum FAIR beam intensities will
be used and interaction rates of 10~MHz allow to collect on the
order of $10^{6}$ J/$\psi$. Yields, spectra and first flow
measurements of charm production can be performed with these
statistics. Measuring both, open charm and charmonium production
allows to study the propagation of produced charm quarks in the
dense medium and address the question whether their behaviour is
more quark or (pre-)hadron like.

\begin{figure}[ht]
\centering
\includegraphics[height=0.245\textheight]{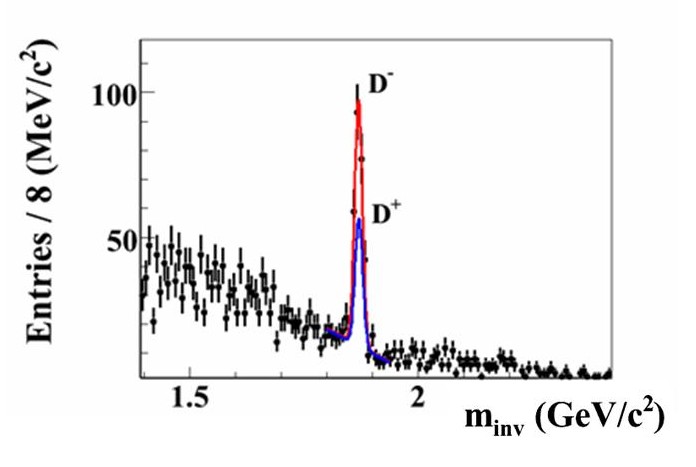}
\hfill
\includegraphics[height=0.245\textheight]{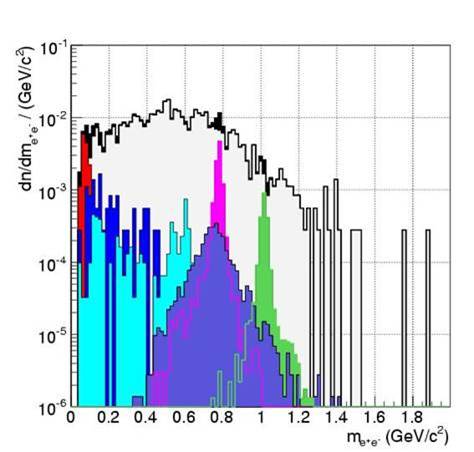}
 \caption{Feasibility studies for the reconstruction of
D$^{\pm}$-mesons (left, $K\pi\pi$ decay channel, $10^{9}$ events)
and low-mass di-electrons (right, different sources from left to
right are $\pi^{0}$-, $\eta$-,
 $\omega$-Dalitz, $\rho$-, $\omega$-, $\phi$-meson, 200k events) in central Au+Au collisions at 25 $A$GeV. Background events are
 taken from UrQMD \cite{urqmd}, signal multiplicities from HSD \cite{hsd2}.}
\label{cbm_figs}
\end{figure}

In order to meet the experimental challenge the CBM detector is
designed for hadron and lepton identification, and for high
precision tracking and secondary vertex resolution \cite{cbm}.
Right behind the target and inside a large dipole magnet radiation
hard and fast silicon pixel and strip detectors will be placed.
Detectors for particle identification will follow downstream. Two
different setups will allow for either electron or muon
identification. The detector development is in process and
detailed feasibility studies investigate the expected performance
and help to optimize the setup. Two examples of key observables
for CBM are shown in figure \ref{cbm_figs}.

\section{Conclusion and Outlook}

The SPS low-energy program has been a very successful endeavour:
Strong evidence has been found for the onset of deconfinement at
30 $A$GeV beam energy. This conclusion is based on a variety of
measurements showing distinct changes in the energy dependence of
yields, spectra and correlations of the emitted particles at this
energy. Modifications in the particle emission due to the initial
phase are clearly seen in di-lepton spectra and particle
production at high $p_{t}$. However the data do not allow for a
more quantitative characterization of the created media.

Obviously, the SPS experiments investigated a very interesting
region of the QCD phase diagram in which also structures such as a
first order phase transition between hadronic and partonic matter
and the critical endpoint are predicted. Future investigations
with modern, 2nd generation experiments are therefore planned in
order to find these structures and characterize the created
strongly interacting matter. In this context the planned CBM
experiment at FAIR is of special interest as it will allow to
access also rare probes such as charm production at threshold and
di-leptons due to the high intensity beams provided by FAIR. The
systematic, high statistics and high precision investigations of
all these upcoming experiments will hopefully finally characterize
the intermediate range of the QCD phase diagram within the next
10-20 years.


\section*{Acknowledgments} 
Thanks to the QM09 organizers for the invitation to this review,
and for stimulating discussions and support to A.~Andronic,
H.~Appelsh\"auser, C.~Blume, P.~Braun-Munzinger, M.~Gadzdicki,
F.~P\"uhlhofer, P.~Senger, P.~Seyboth, R.~Stock, J.~Stroth, and
the colleagues in the NA49 and CBM collaborations.

\end{document}